\documentclass[11pt]{article}

\usepackage{url} 
\usepackage{graphicx,amssymb,amsmath,amssymb,textcomp,url,booktabs,setspace}
\usepackage[left=1.5in,right=1.5in,bottom=1.5in,top=1.5in]{geometry}
\usepackage[numbers,sort&compress]{natbib}
\usepackage[T1]{fontenc}
\usepackage{ae,aecompl} 

\title{Using causal models to distinguish between neurogenesis-dependent and -independent effects on behaviour}

\author{\textit{Stanley E. Lazic}  \\
 \textit{Bioinformatics and Exploratory Data Analysis, F. Hoffmann-La Roche} \\
  \textit{4070 Basel, Switzerland} \\
  \url{stan.lazic@cantab.net}
	}

\date{}

\begin{document}

\maketitle

\begin{abstract}
  There has been a substantial amount of research on the relationship between hippocampal neurogenesis and behaviour over the past fifteen years, but the causal role that new neurons have on cognitive and affective behavioural tasks is still far from clear. This is partly due to the difficulty of manipulating levels of neurogenesis without inducing off-target effects, which might also influence behaviour. In addition, the analytical methods typically used do not directly test whether neurogenesis mediates the effect of an intervention on behaviour. Previous studies may have incorrectly attributed changes in behavioural performance to neurogenesis because the role of known (or unknown) neurogenesis-independent mechanisms were not formally taken into consideration during the analysis. Causal models can tease apart complex causal relationships and were used to demonstrate that the effect of exercise on pattern separation is via neurogenesis-independent mechanisms. Many studies in the neurogenesis literature would benefit from the use of statistical methods that can separate neurogenesis-dependent from neurogenesis-independent effects on behaviour.
\end{abstract}


\clearpage

\section{Introduction}
 One of the advantages of laboratory-based studies is that it is often possible to experimentally control important variables that might influence the system under investigation. This is a luxury that some fields of biology (e.g. ecology) and many social sciences do not have. Instead, they often have to measure these relevant variables and use more complex statistical models to understand cause-and-effect relationships. Occasionally, some aspect of the system is not under complete experimental control, and it is necessary to use analytical methods developed in other fields to address key research questions. Such methods are often necessary when relating adult hippocampal neurogenesis to behaviour because factors that influence neurogenesis such as age, hormones, stress, physical activity, environmental enrichment, anti-depressants, disease status, etc. have off-target effects, and therefore establishing a causal relationship between neurogenesis and behaviour is not straightforward. Methods have been developed to test whether hypothesised causal relationships such as $A$ affects $B$, and $B$ affects $C$ (but $A$ does not affect $C$ directly)  are consistent with the data (denoted as $A \rightarrow B \rightarrow C$). Here, $B$ is said to mediate the effect of $A$ on $C$, and thus $A$'s effect on $C$ is indirect. However, a direct $A \rightarrow C$ effect might exist that does not involve $B$. How would one distinguish between these two situations? Or can $A$ have an effect on $C$ both via $B$ and directly? If so, what proportion of the effect is via $B$? Causal models can be used to address these types of questions, but are rarely employed in studies of adult hippocampal neurogenesis. This means that the analyses do not directly address the question that the study was designed to test, and incorrect conclusions can be reached regarding how $A$ affects $C$ (i.e. directly, via $B$, or both). 
 
	In a recent article,  Creer et al. \cite{Creer2010} examined the relationship between neurogenesis and performance on a pattern separation task. This task had two outcomes that were measured: the number of trials required to reach a set criterion, and the number of reversals achieved after a fixed number of attempts. Levels of neurogenesis were manipulated by placing a running wheel in the cages of some animals (runners), while controls had no running wheel. They showed that there was a relationship between neurogenesis and behaviour for the reversal data, and a non-significant trend in the predicted direction for the trials-to-criterion data.  They concluded that because exercise (1) increased levels of neurogenesis, and (2) led to better performance on the behavioural task, that neurogenesis might be important for this task. However, this conclusion does not follow from the analysis because the effect of exercise on other aspects of neural functioning have not been taken into consideration. In other words, they concluded that $A$ (exercise) affects $B$ (neurogenesis), which affects $C$ (behaviour). What about the direct $A \rightarrow C$ effect, which subsumes all of the neurogenesis independent effects? Could this completely account for the observed results?

	Figure 1 shows two potential models of the causal relationships (indicated by arrows) between the treatment, neurogenesis, and behaviour.  In the first model, the treatment (exercise vs. control) influences neurogenesis, which in turn influences behaviour (performance on a pattern separation task). In the second model, neurogenesis does not have a direct causal effect on behavioural performance. The treatment influences neurogenesis, and it influences behaviour directly (or more likely via some other variable that was not measured, such as changes in the electrophysiological properties of the cells, spine density, etc). These two models make different predictions about relationships that will be found in the data, and therefore these predictions can be tested against the data. The model that correctly predicts the data will have the greater support, and as will be shown below, the evidence is strongly in favour of the neurogenesis-independent model. In addition to the two different predictions, these models also make three predictions which in common, and therefore these three predictions cannot be used to discriminate between the models.  Unfortunately, many published studies only test the common predictions, and therefore they cannot provide support for a causal role for neurogenesis.

\section{Methods}
 Data were accurately extracted from Figure 2B and 3C of the original publication \cite{Creer2010} using g3data software (\url{www.frantz.fi/software/g3data.php}). This reanalysis focuses mainly on the reversal data, since it had a significant relationship with neurogenesis.  Analysis was conducted with R (version 2.12.2; \cite{Ihaka1996,R2011}). In order to confirm the accuracy of the data extraction, the same analysis as Creer et al., was performed.  R$^2$ between neurogenesis and behavioural performance was determined to be 0.235, which compares favourably with the results reported in the original publication (R$^2$ = 0.236; percent error $<$ 0.5\%). The data and R code are provided in the supplementary material so that the analysis is documented and reproducible.

 Two methods were used in the reanalysis of the data. The first is from Baron and Kenny \cite{Baron1986}, which can be implemented with standard software, does not require  advanced statistical knowledge, and is even discussed in introductory statistics textbooks \cite{Judd2009}. More recent developments and caveats of using this approach are discussed in references \cite{shrout2002,Mackinnon2002,MacKinnon2007,Geneletti2007,Kraemer2008,Mackinnon2009,Hayes2009,VanderWeele2009,Zhao2010}. In this simple experiment (only three variables with linear relationships) the relevant questions can be addressed with t-tests, linear regression, and an analysis of covariance (ANCOVA)---all standard techniques that will be familiar to many biologists. An ANCOVA model is characterised by a continuous response variable with a combination of continuous and categorical predictor variables. In this case, performance on the pattern separation task was the response variable, neurogenesis was the continuous predictor, and treatment (exercise vs. control) was the categorical factor. In the output of such an analysis, the effect of neurogenesis is adjusted for the effect of the treatment, and the effect of the treatment is adjusted for the effect of neurogenesis, and these results are conveniently the two predictions which differ between the models. These tests are all just specific examples of a linear model, and all of the above tests can be conducted within a linear modelling framework, which is a more unified method for data analysis \cite{Cohen1968,Judd2009,Rodgers2010}. All of the results are reported as regression coefficients from linear models, but the R code in the supplementary material contains the equivalent results presented as t-tests, regression, and ANCOVA.  The key relationships are described with the following equations, where $T$, $N$, and $B$ represent the variables Treatment, Neurogenesis and Behaviour, respectively.

 \begin{eqnarray}
      N & = & \alpha_1 + \beta_1T + \epsilon_1 \\
      B & = & \alpha_2 + \beta_2T + \epsilon_2 \\
      B & = & \alpha_3 + \beta_3N + \epsilon_3 \\
      B & = & \alpha_4 + \beta_4T + \beta_5N + \epsilon_4 
   \label{eq:BandK}   
 \end{eqnarray}

\noindent Equations 1--3 test model predictions 1--3 in Figure 1, and Equation 4 tests model predictions 4 and 5. The $\alpha$ parameters represent the intercept for each model and are not of direct interest. The $\beta$ parameters are the coefficients for the effect in question and the subscript corresponds to the model prediction that is being tested. The $\epsilon$'s are the error terms (residuals).  In the context of this experiment, the main interest is in the relationship between neurogenesis and behaviour, once the effect of exercise on behaviour has been removed (i.e. testing whether $\beta_5$ is different from zero). A related question is whether the effect of exercise is mediated via neurogenesis (for example, if the main research question is ``By what mechanism does exercise affect behaviour''). This can be determined by requiring that (1) the treatment affects neurogenesis, and (2) neurogenesis is associated with behaviour after adjusting for a direct effect of treatment on behaviour (i.e. $\beta_1$ and $\beta_5$ both need to be significant). However, this piecemeal approach is not ideal because two statistical tests are performed and the strength of the relationship is not quantified. An alternative is the product-of-coefficients method where $\beta_1 \times \beta_5$ needs to be significantly different from zero. One difficulty with this approach is estimating the standard error, or uncertainty, of the estimate \cite{Mackinnon2002,shrout2002}; however, this can be easily estimated in the Bayesian analysis below.

	The second method uses a Bayesian graphical model to estimate causal relationships. This is a general method, which can be scaled up to more complex experimental designs,  provide more robust estimates with small samples sizes, and allows latent variables and prior information from previous experiments to be included. In addition, direct probability statements can be made about the parameters, and the results can provide support for the null hypothesis \cite{Gallistel2009}; traditional null hypothesis significance testing can only reject (or fail to reject) the null hypothesis, but cannot directly provide support for it. Graphical models are an active area of research with developments occurring in many fields, and as often happens, similar methods are developed with different names (e.g. structural equation modelling, path analysis, Bayesian networks, probabilistic graphical models, causal mediation analysis), along with slightly different philosophies, emphases, assumptions, and underlying algorithms  \cite{Shipley2000,Grace2006,Lee2007,Kjaerulff2008,Pearl2009,Yuan2009,Imai2010}. A general introduction to probabilistic graphical models using gene expression and cell signalling as examples are provided by Needham et al. \cite{Needham2006,Needham2007} and Friedman \cite{Friedman2004}, and a more detailed description of these methods using a simple three-variable problem can be found in Yuan and MacKinnon \cite{Yuan2009}. Briefly, the hypothesised relationships are written as directed acyclic graphs (DAGs; e.g. Fig. 1), where each variable is represented as a node, and hypothesised causal relationships are represented by arrows. These graphical representations are then converted into a set of equations (similar to Equations 1 and 4). The joint distribution of all of the variables are factored into a set of simpler conditional distributions, much like the number 12 can be factored into $3 \times 2 \times 2$. Graphical models with a different arrangement of hypothesised causal relationships will be factored in different ways (e.g. 12 can also be factored into $6 \times 2$ and $4 \times 3$), following a few simple rules. If a variable has no arrows pointing into it (i.e. it is not dependent on another variable), then it is written as a regular probability density or distribution function (e.g. $P(X)$ is the probability density function of a continuous variable $X$). Variables which have an arrow pointing into them are condition on those variables. For example, if there is an arrow pointing from $Y$ to $X$, this would be written as $P(X|Y)$, which is read as the probability of $X$ given $Y$ (the ``|'' means ``given''). Thus, the joint probability of the three variables $P(T,N,B)$ can be rewritten as a set of conditional distributions, which correspond to the two graphical models (Fig. 1). For Model 1, this can be written as 

        \begin{equation}
          P(T,N,B) = P(B|N)P(N|T)P(T),
        \end{equation}

\noindent whereas Model 2 would be written as

        \begin{equation}
        P(T,N,B) = P(B|T)P(N|T)P(T),
        \end{equation}
  
\noindent with the only difference being whether $B$ is conditioned on $N$ or $T$, reflecting the two potential variables that affect behaviour. It is then possible to test whether there is any evidence in the data for the hypothesised causal structures represented by these equations, and to quantify the degree to which one model is better compared to the other. 

Analysis was conducted with the R2WinBUGS R package \cite{Sturtz2005} and OpenBUGS, and the code is provided in the supplementary material. Non-informative priors were used and the results were not sensitive to the form of the prior (e.g. normal versus uniform). Three chains with 500,000 iterations each were used for the Markov-chain Monte Carlo (MCMC) sampling, with a burn in period of 250,000 iterations, and every tenth value was saved. The three chains were well mixed (Gelman-Rubin statistic $< 1.01$ for all parameters). The main parameter of interest was the effect of neurogenesis on behaviour after adjusting for the indirect effect of exercise on behaviour ($\beta_5$). In addition, the direct ($\beta_4$) and indirect ($\beta_1 \times \beta_5$) effect of exercise on behaviour was determined.

Many of the requirements for using causal models are the same as for a standard analysis, including a large enough sample size to obtain reasonably precise estimates, important variables have not been omitted, random assignment of animals to groups, and randomised collection of data, processing of samples, and quantification to avoid time-dependent confounders. Causal models have a few additional assumptions. First, the mediator (i.e. neurogenesis) should ideally be measured without error \cite{Baron1986}. Creer et al. expressed the neurogenesis data as a density measurement (neurons/mm$^3$), and so the extent to which values represent changes in cell number (which is of interest) or changes in the volume of the dentate gyrus is not clear. It is assumed that the reported estimates are an accurate representation of the number of cells and are not influenced by changes in volume between conditions. However, such assumptions can be misleading and are unnecessary when using modern design-based stereological methods \cite{Oorschot1994,Mouton2002,Mayhew2003}. Second, there should be no neurogenesis by treatment interaction, which was reasonable to assume with this data (F(1,16) = 0.17, p = 0.687), although more recent methods can handle interactions \cite{VanderWeele2009}. Third, the residuals $\epsilon_1$ and $\epsilon_4$ should not be correlated, and levels of neurogenesis and $\epsilon_4$ should not be correlated, as this might indicate that another variable not included in the analysis is affecting levels of neurogenesis and behaviour (both correlations were almost zero $p > 0.999$). Finally, it is assumed that the arrows are pointing in the right direction. This may be difficult to check in general, but because this was a randomised experiment, we can be certain that levels of neurogenesis and behavioural performance did not affect assignment to different treatment groups.

\section{Results}
\subsection{Predictions common to both models}
The first prediction common to both models is that levels of neurogenesis will differ between the runners and control mice. This is indeed the case, with mice in the exercise condition having approximately twice as many new cells (Fig 2A, $\beta_1 = 3.23$, 95\% CI =  2.03 to 4.43, p $<$ 0.001).

	The second prediction common to both models is that performance on the behavioural task will differ between the runners and control mice.  The neurogenesis-dependent model predicts this because the treatment influences neurogenesis, and levels of neurogenesis in turn influence performance on the pattern separation task. The neurogenesis-independent model predicts this because exercise influences pattern separation directly (or more likely via some unmeasured variable).  The runners had more than twice as many reversals compared to the control group (Fig 2B, $\beta_2 = 0.96$, 95\% CI =  0.40 to 1.52, p = 0.002), which is consistent with both models. 

	The third prediction common to both models is that there will be a relationship between neurogenesis and behaviour.  The neurogenesis-dependent model predicts this because neurogenesis has a causal influence on pattern separation. This is the main research question that the study was designed to test. The problem is that the neurogenesis-independent model makes the exact same prediction.  This is because neurogenesis and behaviour are influenced by the same variable (the treatment) which induces a correlation between them. This is the classical third-variable problem, where A and B are correlated, but only because they share a common cause in variable C. The relationship between neurogenesis and behaviour is significant (slope of the solid line in Fig 2B: $\beta_3$ = 0.18, 95\% CI = 0.02 to 0.34, p = 0.030), and Creer et al. took this as evidence that neurogenesis might be involved in the pattern separation task. However, since the neurogenesis-independent model makes the same prediction, further analyses are required to separate the neurogenesis-dependent from the neurogenesis-independent effects.

\subsection{Predictions that discriminate between models}
	If neurogenesis is causally involved in the pattern separation task, then blocking the increase in neurogenesis caused by exercise will cause the performance of the runners and controls to be the same (that is, the relationship between treatment and behaviour will no longer be significant). We could imagine doing this experimentally, where the mice would be given a compound whose sole effect is to nullify the effect of exercise on neurogenesis (indeed, this approach is often taken, and studies with this design are discussed below).  Exercise would not affect behaviour in this case because the causal link between treatment and behaviour has been broken by holding neurogenesis at a constant level. The neurogenesis-independent model however predicts that the relationship between treatment and behaviour will still exist after keeping the levels of neurogenesis constant. This is clear from Figure 1; in the neurogenesis-independent model there is a direct causal link between the treatment and behaviour, so whether levels of neurogenesis are held fixed or not is irrelevant. It is also possible to statistically hold constant, or fix, levels of neurogenesis at some value (also referred to as ``adjusting for'', ``taking into account'', ``controlling for'', or ``conditioning on''), and when this is done we find that there is still a significant relationship between treatment and behaviour (difference between runners and controls: $\beta_4$ = 1.06 reversals, 95\% CI = 0.09 to 2.02, p = 0.034), thus providing support for the neurogenesis-independent model. 

	The other prediction that differs between the models (prediction five in Fig. 1) is that if the neurogenesis-independent model is correct, then conditioning on the treatment will remove the correlation between neurogenesis and behaviour. This is clear from Figure 1, where it can be seen that in the neurogenesis-independent model, the link between neurogenesis and behaviour will be broken if we condition on the treatment. This can be thought of as removing a common cause of neurogenesis and behaviour, and testing whether there is still a relationship between them. However, the neurogenesis-dependent model predicts that there will still be a relationship between neurogenesis and behaviour because there is a direct causal link between the two. The results show that there is no relationship between neurogenesis and behaviour when conditioned on treatment ($\beta_5$ = -0.03, 95\% CI = -0.27 to 0.21, p = 0.788, represented as the slope of the two dashed lines in Figure 2C), which also supports the neurogenesis-independent model. Note that this is not due to a lack of statistical power, the value is close to zero, and is even in the opposite direction to what the neurogenesis-dependent model predicts (i.e. higher levels of neurogenesis were associated with worse performance, although not significantly so). This model is actually fitting a separate regression line for the runners and controls, rather than one regression line through all of the data points as in Figure 2C. The relationship between neurogenesis and behaviour needs to hold within each group, and should not be driven by differences between the group means of the two variables.

The above analysis examined individual parameters to see whether they were different from zero, which is useful to address particular questions about the data. It is also possible to examine a model as a whole, by calculating a measure of model fit or adequacy. Two models can be fit to the same data and then compared, and the model with the better fit is preferred. Models that are more complex (have more parameters) have greater freedom to better approximate the data and therefore will fit better than less complex models. Therefore the complexity of the models must also be taken into account when performing a comparison (in this case the two models had the same number of parameters). A number of methods have been developed for this type of model comparison, and the deviance information criterion (DIC; \cite{Spiegelhalter2002}) is one appropriate option for the Bayesian analysis. The lower the DIC for a model, the better the fit, and the larger the difference in DICs between two models, the better one model is versus the other. There are no strict rules regarding how large a difference is considered important, however a difference in DIC between 5--10 can be considered substantial   (\url{http://www.mrc-bsu.cam.ac.uk/bugs/winbugs/dicpage.shtml}). DIC for the neurogenesis-dependent model was 116.3, while for the neurogenesis-independent model it was 110.9. The difference of 5.4 therefore provides strong support for the neurogenesis-independent model.

The above analysis pitted two competing models against each other that hypothesised different mechanisms for the effect of exercise on behaviour. While this is useful for explanatory purposes, it is likely that multiple mechanisms are at work (both neurogenesis and neurogenesis-independent), and the main research question should be ``\textit{to what extent} does neurogenesis contribute to behaviour''. In other words, it would be useful to determine how much of exercise's effect is due to neurogenesis, and how much is due to neurogenesis-independent mechanisms. The relationship between exercise and behaviour was therefore decomposed into the indirect/neurogenesis-dependent ($Exercise \rightarrow Neurogenesis \rightarrow Behaviour$) and direct/neurogenesis-independent ($Exercise \rightarrow Behaviour$) paths, which were then estimated. This can be represented as

        \begin{equation}
          P(T,N,B) = P(B|T,N)P(N|T)P(T),
        \end{equation}

\noindent where the first term now has $B$ conditioned on both $T$ and $N$ (compare with Equations 5 and 6). The results are displayed in Figure 2D, which show the posterior densities for the two effects. These density plots represent the estimated effects along with the uncertainty associated with them. For example, the direct effect (DE) of exercise via neurogenesis-independent mechanisms is to increase the number of reversals by approximately one (the peak of the distribution), with a 95\% credible interval of 0.09 to 2.03 reversals (a 95\% credible interval means that there is a 95\% chance that the true effect is between the upper and lower values, this is often what confidence intervals are mistakenly believed to be). Since the lower interval excludes zero, we can conclude that there are neurogenesis-independent mechanisms at work (the associated p-value is 0.016). However, the indirect effect (IE) of exercise via neurogenesis on behaviour was very close to zero, and was slightly negative. In other words, the ``best guess'' of the effect of increasing neurogenesis is that it worsens performance (which is the same result as the first analysis). If the effect of neurogenesis is actually zero, a negative value would be expected 50\% of the time, so this is not surprising. It is clear from these results that exercise is not affecting behaviour via changes in levels of neurogenesis.

	The results are the same for the trials-to-criterion data. Creer et al. established that the first two predictions are supported by the data. The third prediction (a relationship between neurogenesis and behaviour) was not significant ($\beta_3$ = -3.79, 95\% CI = -8.74 to 1.15, p = 0.125), but they described it as a ``trend''. However, this relationship between neurogenesis and behaviour is greatly attenuated ($\beta_5$ = -0.16, 95\% CI = -8.37 to 8.04, p = 0.967) when conditioned on treatment, indicating that there is little support for the neurogenesis-dependent model.

	Creer et al., concluded that ``Taken together, our findings indicate that exercise-induced neurogenesis improves dentate gyrus-mediated encoding of distinct spatial representations'', and it seems that this conclusion has been accepted by others \cite{Tronel2010,Oomen2011,Klaus2011,Inokuchi2011,Yassa2011}. However, a reanalysis of their data using causal models---which can tease apart complex relationships and directly test whether neurogenesis mediates the effects of exercise on behaviour---clearly shows that the effect of exercise is via neurogenesis-independent mechanism(s).  This is further supported by their finding in a separate experiment where aged runners (the reanalysed experiment used younger adults) did not have an increase in neurogenesis but still showed a small improvement on a modified version of the pattern separation task, which is exactly what the neurogenesis-independent model predicts. It should also be noted that the adult mice ran 23.5 $\pm$1.79 km/day while the aged mice only ran 5.4 $\pm$0.68 km/day. Therefore, because aged animals ran 77\% less, the effect of running (via neurogenesis-independent mechanisms) will be predicted to influence behaviour to a lesser extent. While this manuscript was under review, this group has published a review article discussing other factors that may mediate the effects of exercise and environmental enrichment on behaviour \cite{Bekinschtein2011}. We need to go further however, and convert these pictorial representations of hypothesised causal relationships into statistical models which can be tested against the data.

\section{Discussion}
Causal models are rarely used to analyse data from laboratory-based biological experiments, which is partly due to the high experimental control that can usually be achieved in such situations (lack of knowledge of these methods is another reason). However, it is not always possible to create two experimental groups that differ only on the variable of interest. For example, the runners are not the same as the controls in all respects (even though they may have been at baseline due to randomisation), with the only exception of having higher levels of neurogenesis. If these other variables are potential candidates for mediating the effect of the treatment, then it is not possible to make causal statements about the variable of interest. One could imagine the exact same study (and analysis) carried out by a second group whose interest is not in neurogenesis, but in spine density (also affected by exercise \cite{Stranahan2007,Eadie2005}), which was measured instead. Another group would look at levels of synaptic proteins, while a fourth would focus on glutamate receptors (both of which are also affected by exercise \cite{Hu2009,Farmer2004}). Without the appropriate analysis, each group would conclude that their hypothesised variable is important for pattern separation, and all four would only have an incomplete picture. It is possible that a particular variable has no effect, and thus the wrong conclusion would be reached. If more than one variable is important, than all of the estimated effects will be biased because the total effect (via all mechanisms) is being estimated, rather than the individual effects separately.

While there was no evidence for a causal role of neurogenesis in the present experiment, this does not rule out a contribution of neurogenesis to other behavioural tasks, or for the effect of decreasing neurogenesis on pattern separation \cite{Clelland2009}. Indeed, the causal relationships are likely to be more complex than suggested by this analysis, since angiogenesis was also affected by the treatment, and this introduces another variable by which the treatment might affect behaviour \cite{Lazic2010a}.

\subsection{Manipulating neurogenesis often affects other relevant variables}
        If the goal of a study is only to demonstrate the effect of exercise on behaviour, then the concerns discussed herein do not apply. However, problems arise if the interest is in the effect of neurogenesis on behaviour, and exercise (or some other treatment) is used as a means to manipulate neurogenesis. Most methods of manipulating neurogenesis also affect other variables that are known (or could reasonably be expected) to contribute to behaviour (Table 1). For example, exercise affects spine density in the entorhinal cortex, CA1, and dentate gyrus \cite{Stranahan2007,Eadie2005}, as well as levels of synaptic proteins \cite{Hu2009}, glutamate receptors \cite{Farmer2004}, and various other genes, growth factors, and neurotransmitters in the dentate gyrus \cite{Molteni2002,Lista2010}. If other factors are known to be involved, it is not possible to address the question ``what is the effect of neurogenesis on behaviour'' simply by looking at the relationship between the two. This will lead to severely biased results if the other relevant factors are not taken into account, as was demonstrated in the present study. Once it is known that exercise affects other aspects of the brain which might provide alternative causal explanations, future studies must take them into account, or at least rule them out as a potential contributing factor in any particular study. In addition to experimental manipulation, some studies use groups that naturally vary in levels of neurogenesis such as young versus old, or a disease model versus control. Here, neurogenesis is not manipulated directly, but clearly there are many differences between old and young brains, or in transgenic/knock-out mice versus controls that need to be ruled out. Importantly, anything that affects physical activity or locomotor behaviour in the home cage (e.g. a disease model, general health, age) might then influence performance on a behavioural task.

This reanalysis not only reverses the conclusions of the original study, but brings into question the conclusion of many published studies purporting to demonstrate evidence of an association between neurogenesis and behaviour. This is because the basic design, analysis, and logic of the Creer et al. study is similar to many studies in the literature: manipulate neurogenesis and observe behaviour; if there are differences in both neurogenesis and behaviour between groups, conclude (tentatively) that neurogenesis might be causally involved. Furthermore, ignore other potential variables that might explain the results. The details of the studies vary (e.g. method of manipulating neurogenesis, type of behavioural task, etc.), but they follow the same basic template.  The onus is on those reporting associations between neurogenesis and behaviour to demonstrate that neurogenesis-independent mechanisms were not at work, or at least quantify their magnitude and estimate the unique contribution that neurogenesis makes. For many studies this has not been done, and there is good reason to believe that neurogenesis-independent effects play a role in these studies (see Table 1 for a brief list). It is possible (and indeed not difficult) to design studies with \textit{biased} effects that can be consistently reproduced \cite{Rosenbaum2001}, and it is likely that this has occurred in the neurogenesis literature. The possibility that neurogenesis-independent effects might play a role is acknowledged in many studies and review articles (e.g. Creer et al. noted that exercise is known to affect expression of neurotrophins, vascularisation, dendritic spine density, and synaptic plasticity), but the functional consequences of this when analysing the data and interpreting the results are generally ignored. If these off-target effects are not under experimental control, then the only option is to use more complex statistical models to account for them.

Many studies only establish that the first two predictions in Figure 1 hold, and some studies such as Creer et al. test the third prediction, but given that these results are also entirely consistent with a system in which the behavioural task is completely independent of neurogenesis, these results cannot support the hypothesis that hippocampal neurogenesis influences behaviour. Therefore statements such as ``Thus, the positive correlation between enhanced neurogenesis and improved learning\ldots\ remains a valid basis for the suggestion that newly generated cells may be important for memory function.'' \cite{vanPraag2008}, and ``Clearly, the results of the past 15 years of research support the idea of a functional role of adult-generated neurons in learning and memory processes\ldots'' \cite{Brueljungerman2011}, need to be reconsidered, as they have less empirical support than commonly supposed.  It should be noted that these are not poorly designed or flawed studies; the analysis simply does not directly test the main research question, or take into account potential confounding variables, which can lead to incorrect conclusions.  Recent review articles comment on the conflicting results in the literature \cite{Abrous2008,Deng2010,Glasper2011,CastillaOrtega2011,MarinBurgin2011}, and often suggest that this is due to some interaction between the method of manipulating neurogenesis, the method of measuring neurogenesis, the species or strain of animal used, differing demands of the behavioural task, the type of memory examined, etc. While these factors most likely play a role, conflicting results are not unexpected if important factors, often unmeasured, are influencing the results. It would be useful to reanalyse many of the key studies using causal models. These studies only estimated the total effect of their respective treatments on behaviour, and if any part of these effects were via neurogenesis-independent mechanisms, then the estimates were biased and attributed too much of the total effect to neurogenesis. If neurogenesis-independent effects were present, then separating the total effect into a direct (neurogenesis-independent) and indirect (neurogenesis-dependent) component would decrease the estimated influence of neurogenesis. Thus, studies that were initially statistically significant will have attenuated estimates and larger p-values. One can only speculate on the number of studies whose conclusions will have to be revised. To get a feeling for what this number might be, it is interesting to note that only five studies (including this reanalysis) \cite{Merrill2003,Bizon2004,Castro2010,Lazic2010a} used models that conditioned on important variables, and these have either found no association between neurogenesis and behaviour, or found that higher levels of neurogenesis were associated with worse behavioural performance. 

Recently, Sahay et al. published a very thorough paper which used genetic methods to selectively increase the survival of new born neurons \cite{Sahay2011}. They found that animals in the high neurogenesis group were better able to discriminate  between two similar contexts, suggesting improved pattern separation. However, perhaps the most interesting result was that there were no differences between the normal and high neurogenesis groups on the standard behavioural tests that numerous previous studies showed to be sensitive to levels of neurogenesis, including the Morris water maze, novel object recognition, active place avoidance, forced swim test, and novelty-suppressed feeding. This suggests that previous studies might have been incorrectly attributing differences in behavioural performance to neurogenesis, while the actual effects were via neurogenesis-independent mechanisms. It is likely that genetic methods of manipulating hippocampal neurogenesis are ``cleaner'' in that there are fewer off-target effects, however this is an assumption, and one which can be tested with the data at hand. In other words, whether neurogenesis-independent effects exist, and the magnitude of these effects, is an empirical question which can be easily addressed.

\subsection{More complex designs}
It is recognised that neurogenesis-independent factors can play a role, and some studies explicitly state that they used two methods of decreasing neurogenesis to avoid off-target effects \cite{Denny2011}. In addition, some studies \cite{Meshi2006,Surget2008,Wojtowicz2008,David2009,Deng2009,Schloesser2010,Winocur2011} go a step further and try to demonstrate causality by (1) increasing neurogenesis and showing this improves performance on a behavioural task, and then (2) inhibiting this increase in neurogenesis and showing that the behavioural improvement is lost. This is doing experimentally what the causal models do statistically, and can provide stronger evidence for a causal role of neurogenesis. These hypothesised causal relationships are shown with solid arrows in Figure 3. However, it is still possible for neurogenesis-independent mechanisms to completely explain such results. This could occur if the method of increasing neurogenesis directly affects behaviour (e.g. $Exercise \rightarrow Behaviour$ dashed line in Figure 3), and the method of decreasing neurogenesis also directly affects behaviour (e.g. $Corticosterone \rightarrow Behaviour$ dashed line), without neurogenesis playing a role (no $Neurogenesis \rightarrow Behaviour$ arrow in Figure 3). For example, it is possible that  exercise and corticosterone affect behaviour directly, and also affect neurogenesis as a by-product. The relationships that actually exist in the data can be tested, and there is no need to speculate about which model or interpretation of the data is correct. The use of causal models in such a situation provides a test of the assumption that neurogenesis-independent mechanisms are not operating, and if they are, their magnitude can be quantified. It should also be stressed that this requires no further experimentation, these relationships can be tested with the available data. The supplementary material contains the results of a simulation study where it is demonstrated that using the standard analytical methods, it is not possible to distinguish between a situation where the behavioural outcome is completely dependent on neurogenesis from a situation where neurogenesis plays no role. The causal modelling approach can clearly distinguish these two situations. In addition, these graphical representations are intuitive, make the hypothesised relationships explicit, highlight the implied assumptions (e.g. independence assumptions between variables), and can be converted into statistical models which can be used to make inferences. Furthermore, such model building is the way science advances \cite{Glass2007}.

\section{Conclusions}

Nakagawa and Hauber \cite{Nakagawa2011} suggested five statistical methods that should be used more often by neuroscientists (meta-analysis, mixed-effects modelling, multiple imputation, model averaging, and Markov chain Monte Carlo), and to this list we could add a sixth ``M'': mediation analysis. One reason for the extensive research on hippocampal neurogenesis is that it is widely believed to play a role in some cognitive and affective behaviours. If this is true, then manipulating levels of neurogenesis is a logical approach for improving memory and treating depression. However, if neurogenesis does not have a causal role on behaviour, or if the role of neurogenesis is small compared to the neurogenesis-independent mechanisms of a treatment, then resources would be better spent elsewhere. It is necessary to establish whether observed relationships are causal, as well as their magnitude, especially if the ultimate aim is to manipulate the system for therapeutic ends. Simple methods exist and are routinely used in other fields, and there is no reason for their continued omission in the neurogenesis literature.

\section*{Acknowledgements}

The comments and suggestions from four anonymous reviewers are gratefully acknowledged.

\bibliographystyle{vancouver}

\clearpage

\begin{table}
\caption{A selection of off-target effects for commonly used methods of manipulating neurogenesis. Due to these off-target effects, it is not possible to make causal claims regarding the role that neurogenesis has on behaviour by simply examining the relationship between neurogenesis and behaviour---more complex models are needed. \newline}
\centering
\begin{tabular}{lll} \toprule
Affects neurogenesis & Off-target effects & References \\ \midrule
Stress/Corticosterone & Hippocampal volume &  \cite{Bessa2009}   \\
& Dendrites/spines &  \cite{Hajszan2009,Bessa2009}   \\
&  &  \\

Exercise & Spine density & \cite{Stranahan2007,Eadie2005}   \\
 & Synaptic proteins & \cite{Hu2009}  \\
 & Glutamate receptors & \cite{Farmer2004} \\
 & Plasticity genes & \cite{Molteni2002} \\
 & Growth factors & \cite{Lista2010} \\
 &  &  \\
Environmental Enrichment & Dendrites/spines  & \cite{Bindu2007,Goshen2009,Beauquis2010}   \\
 & BDNF & \cite{Kazlauckas2011} \\
 &  & \\
MAM$^*$ & General health & \cite{Dupret2005} \\
 & Locomotor activity & \cite{Dupret2005} \\
&  & \\
Imipramine & Dendrites/synapses  &  \cite{Chen2008a,Bessa2009,Chen2010}  \\
 &  & \\
Fluoxetine & Dendrites/spines &   \cite{Hajszan2005,Bessa2009}   \\
 &  & \\
Irradiation & NMDA receptors &  \cite{Shi2006}   \\
 & Protein expression &  \cite{Okamoto2009,Wu2010}  \\
 & Inflammation/vasculature &  \cite{Belka2001,Schindler2008,Lee2010}  \\
 & DNA damage &\cite{Yang2009}  \\

\bottomrule
$^*$Methylazoxymethanol Acetate
\end{tabular}
\end{table}

\clearpage

\begin{figure}
\centering
\includegraphics[scale=0.5]{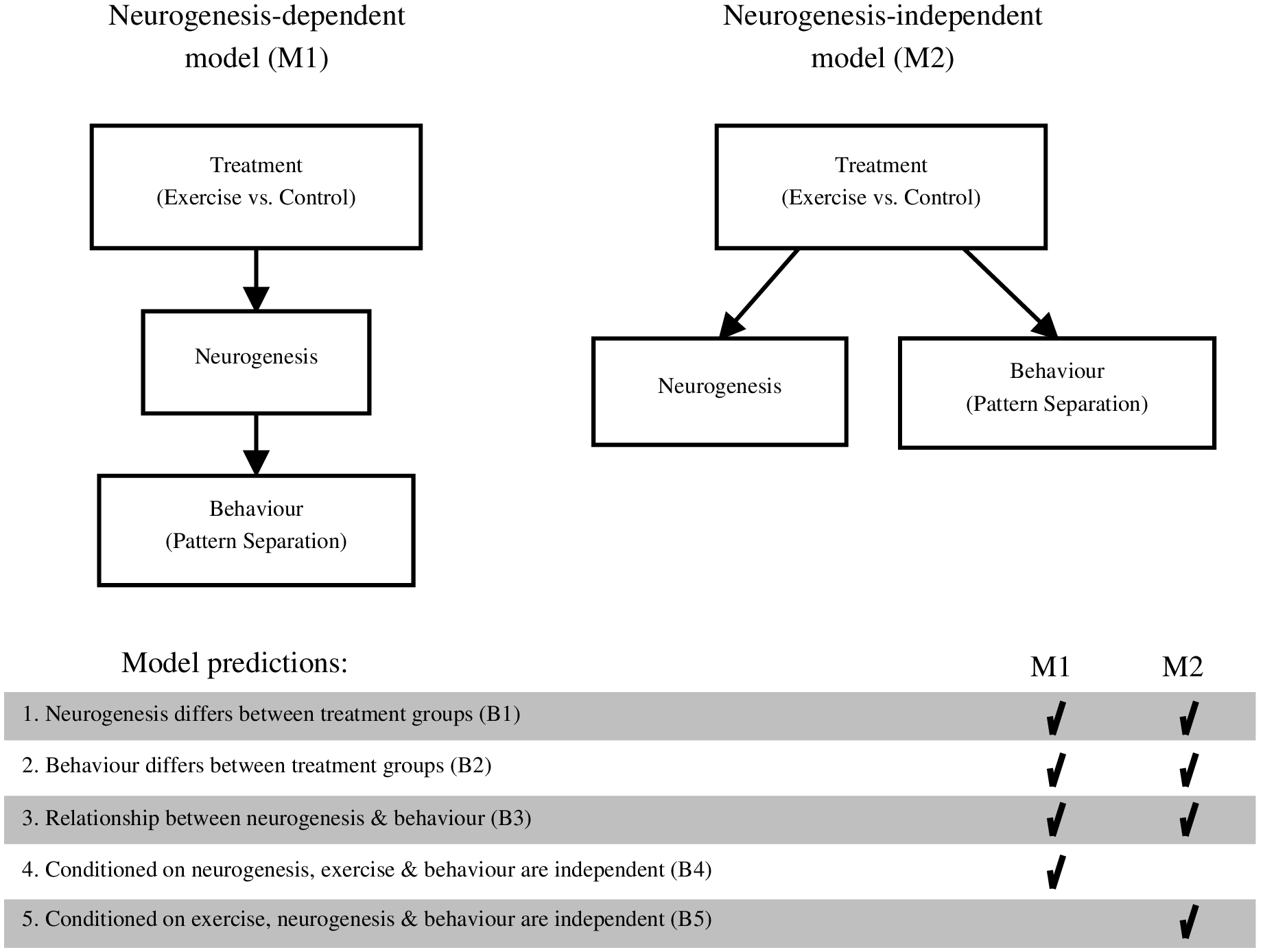}
  \caption{Two hypothetical causal models that explain the relationship between neurogenesis and behaviour. In the neurogenesis-dependent model, an experimental treatment affects the levels of neurogenesis, and altered levels of neurogenesis affect behavioural performance on a pattern separation task. In the neurogenesis-independent model, the experimental treatment affects performance on the behavioural task independently of neurogenesis. These models make three predictions in common and two predictions which differ, and the latter can be tested against the data to provide support for one model versus the other. Arrows indicate hypothesised causal connections. $\beta$'s correspond to the coefficients that are tested in Equations 1--4.}
\end{figure}
\clearpage

\begin{figure}
\centering
\includegraphics[scale=0.8]{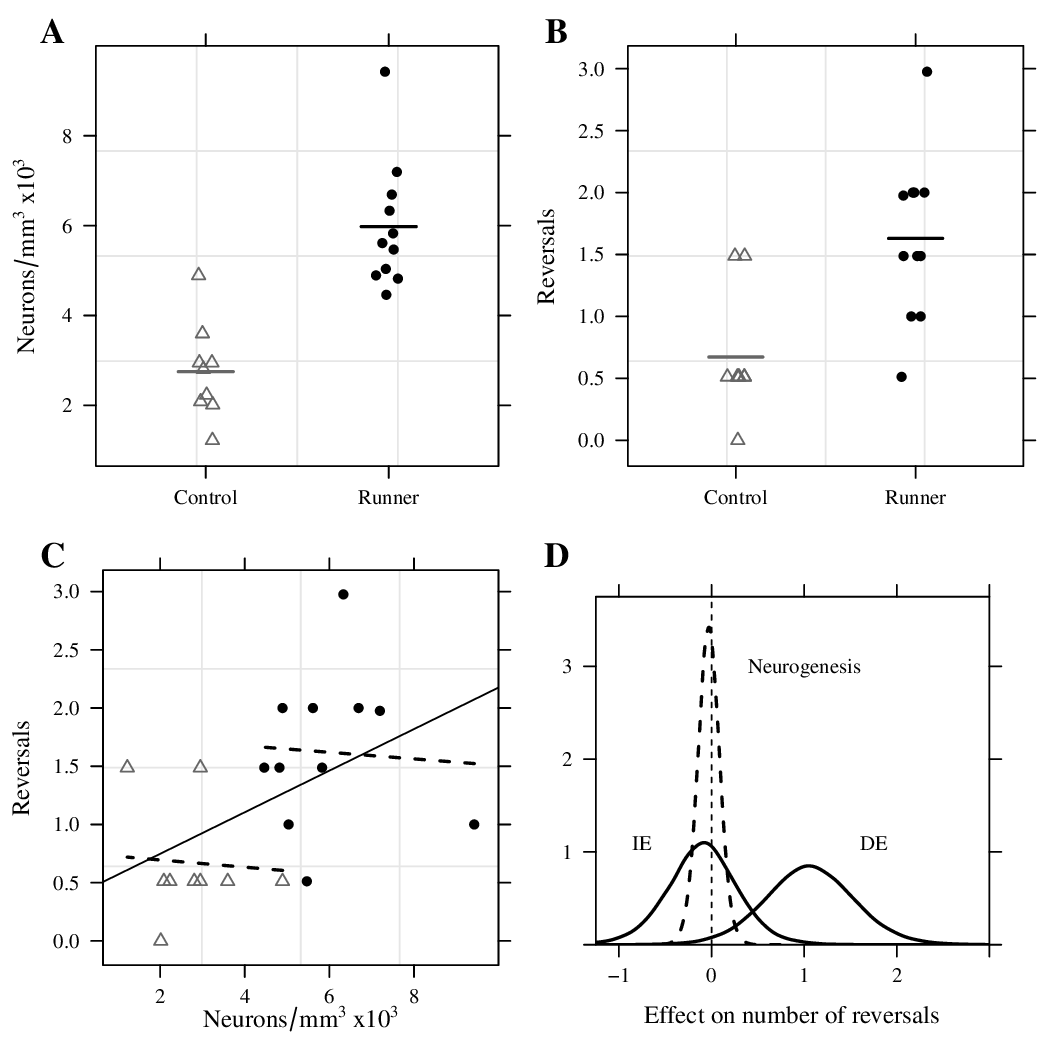}
  \caption{Neurogenesis and behavioural data. Runners had significantly more new cells ($\beta_1$, p $<$ 0.001; A), and more reversals ($\beta_2$, p = 0.002; B) compared to the control group. Most papers in the literature that used a similar design stop the analysis here, incorrectly believing that this provides evidence that neurogenesis affects behaviour. There was also a significant correlation between the two outcomes ($\beta_3$, solid line, p = 0.030, C), but this is still insufficient to indicate that neurogenesis affects behaviour. The dashed lines show the (lack of a) relationship between neurogenesis and behaviour conditioned on treatment group ($\beta_5$). Posterior densities from the Bayesian analysis (D) show that neurogenesis had no effect on the number of reversals conditioned on treatment group ($\beta_5$, dashed line, corresponds to dashed lines in Panel C). Decomposing the total effect of exercise into the indirect effect (IE; via neurogenesis, $\beta_1 \times \beta_5$) and the direct effect (DE; neurogenesis-independent, $\beta_4$) shows that neurogenesis-independent mechanisms are affecting behaviour (p = 0.016). The peak of a distribution is the best estimate of the effect, and the width corresponds to the uncertainty in this estimate.}
\end{figure}
\clearpage

\begin{figure}[ht]
\centering
\includegraphics[scale=0.6]{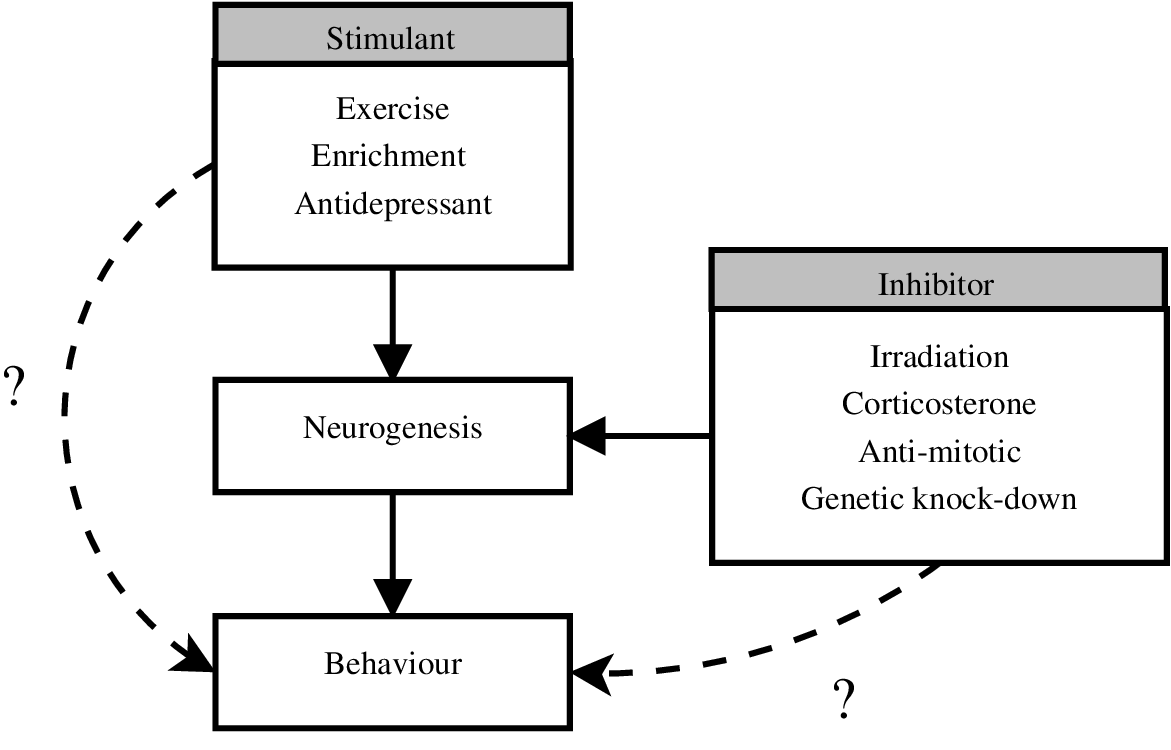}
 \caption{A more complex causal design. Some studies attempt to establish a causal role for neurogenesis by blocking the increase induced by exercise or environmental enrichment using irradiation, anti-mitotic agents, or genetically modified mice. The logic is that if neurogenesis is causally involved in behaviour, then animals in the stimulated$+$inhibited group (i.e. effects of stimulation are reduced or completely blocked) will have reduced behavioural performance compared to animals in the stimulated group. These causal effects are shown with the black solid arrows, with $Neurogenesis \rightarrow Behaviour$ being the main effect of interest. However, this same relationship could be seen if the $Neurogenesis \rightarrow Behaviour$ effect is zero, but the method of stimulating neurogenesis affects behaviour directly ($Stimulant  \rightarrow Behaviour$) and the method of knocking down neurogenesis also negatively affects behaviour directly ($Inhibitor \rightarrow Behaviour$; dashed lines). The important point is that both the initial hypothesised causal relationships as well as the alternatives can be tested against the data. The supplementary material contains a further discussion.}
\end{figure}

\clearpage

\begin{center}
{ \LARGE \textbf{Supplementary material}}
\end{center}

\section*{An examination of ``causal'' experiments}

As mentioned in the main text (Section 4.2 and Figure 3), one method of attempting to establish causality is to manipulate neurogenesis in two ways, first by increasing neurogenesis and looking for a concomitant improvement in behaviour, and then by inhibiting this increase and looking for attenuated improvement on the behavioural task. Such studies can provide much stronger evidence that neurogenesis has a causal influence on behaviour. However, it is still possible that the experimental manipulations affect behaviour completely through neurogenesis-independent mechanisms. Moreover, it is not possible to establish to what extent neurogenesis is involved (if at all) using standard methods of analysis. Simulated data were used to demonstrate this important point. This is required because the causal relationships amongst the variables are known, having been generated by the experimenter. One can then check whether a particular analysis comes to the correct conclusion.  Supplementary Figure 1 shows two causal models from a hypothesised experiment (one neurogenesis-dependent and the other neurogenesis-independent). These graphical models are related to Figure 3 in the main text.

\begin{figure}[ht]
\centering
\includegraphics[scale=0.4]{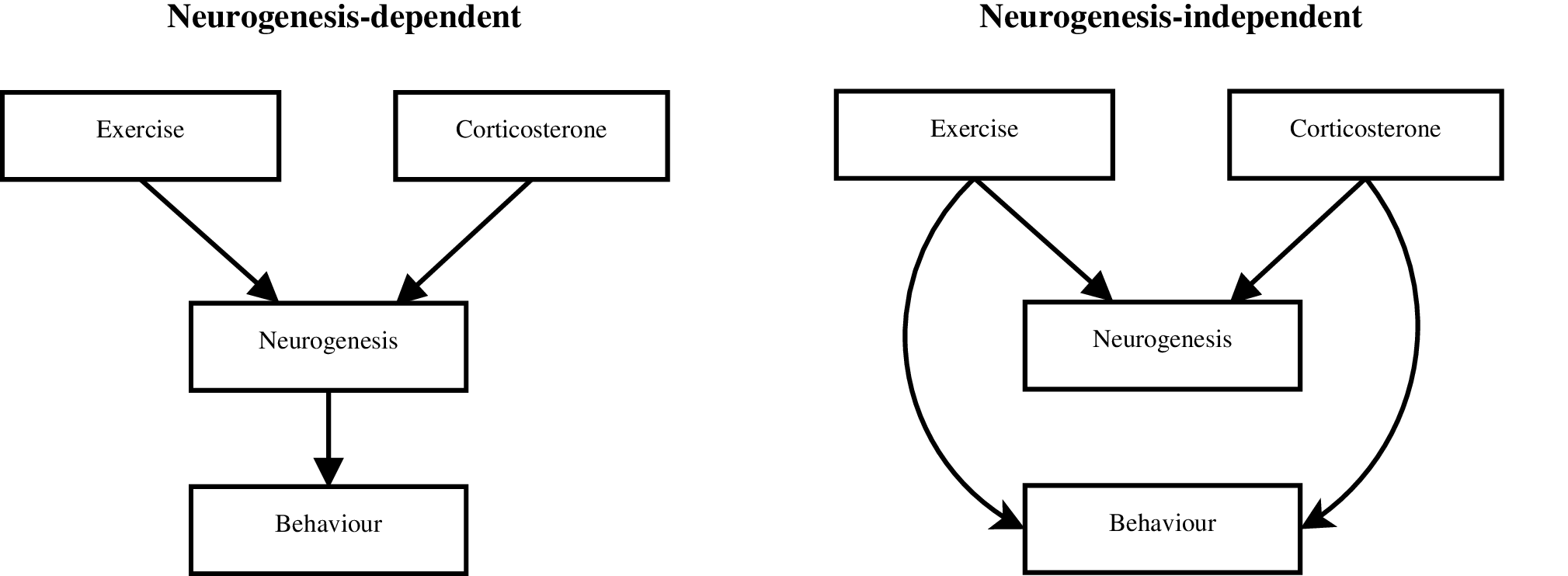}
  \caption{Data were generated from the neurogenesis-independent model and analysed with both standard methods and the causal modelling approach.}
\end{figure}

To make the example concrete, exercise will be the treatment that increases neurogenesis and corticosterone will be the treatment that decreases neurogenesis. The data were generated such that the values and results are similar to what might be found in the literature. Let $E$, $C$, $N$, and $B$ represent the variables Exercise, Corticosterone, Neurogenesis and Behaviour, respectively. Data for the neurogenesis-independent model can be generated using the following equations

\begin{eqnarray}
  N & = & \alpha_1 + \gamma_1E  + \gamma_2C  + \epsilon_1 \\
  B & = & \alpha_2 +  \gamma_1E  + \gamma_2C + \epsilon_2.
\end{eqnarray}

\noindent The $\alpha$ parameters are not of interest and just determine the overall mean value of the data. $\gamma_1$ is the effect of exercise, which effects both neurogenesis and behaviour by ten units. $\gamma_2$ is the effect of corticosterone, which decreases both neurogenesis and behaviour by ten units (the effect of corticosterone is equal and opposite to the effect of exercise and thus they completely cancel each other). The $\gamma$ parameters are known constants for the effect in question and are analogous to the $\beta$ parameters in the main text (which were unknown and estimated from the data). The $\epsilon$'s are the residuals and were drawn from a normal distribution with a mean of zero and a standard deviation of five. The sample size was forty, with ten animals in each condition. Levels of neurogenesis and performance on the behavioural task are in arbitrary units. The following parameter values were used: $\alpha_1 = 25$, $\alpha_2 = 0$, $\gamma_1 = 10$,  $\gamma_2 = -10$,  $\epsilon_1$ and $\epsilon_2 \backsim \mathcal{N}(0,5^2)$. 

For comparison, data from the neurogenesis-dependent model were also generated with the following equations

\begin{eqnarray}
  N & = & \alpha_1 + \gamma_1E  + \gamma_2C  + \epsilon_1 \\
  B & = & \alpha_2 + \gamma_3N + \epsilon_2
\end{eqnarray}

\noindent The same parameter values as above were used, and  $\gamma_3$ is the effect of neurogenesis on behaviour, where a one unit increase in neurogenesis is associated with a one unit improvement in behavioural performance.  Note that Equations 1 and 3 are identical, since neurogenesis is affected by exercise and corticosterone in both models. Equations 2 and 4 differ depending on the model; in Equation 4, behaviour is only affected by neurogenesis, whereas in Equation 2, behaviour affected by exercise and corticosterone, reflecting the causal relationships shown in Supplementary Figure 1. Supplementary Figure 2 plots the generated data from these two models using a bar graph with error bars representing $\pm$1 standard error of the mean (SEM)---the standard method of plotting such data, along with a scatterplot of neurogenesis vs. behaviour.

\begin{figure}
\centering
\includegraphics[scale=0.5]{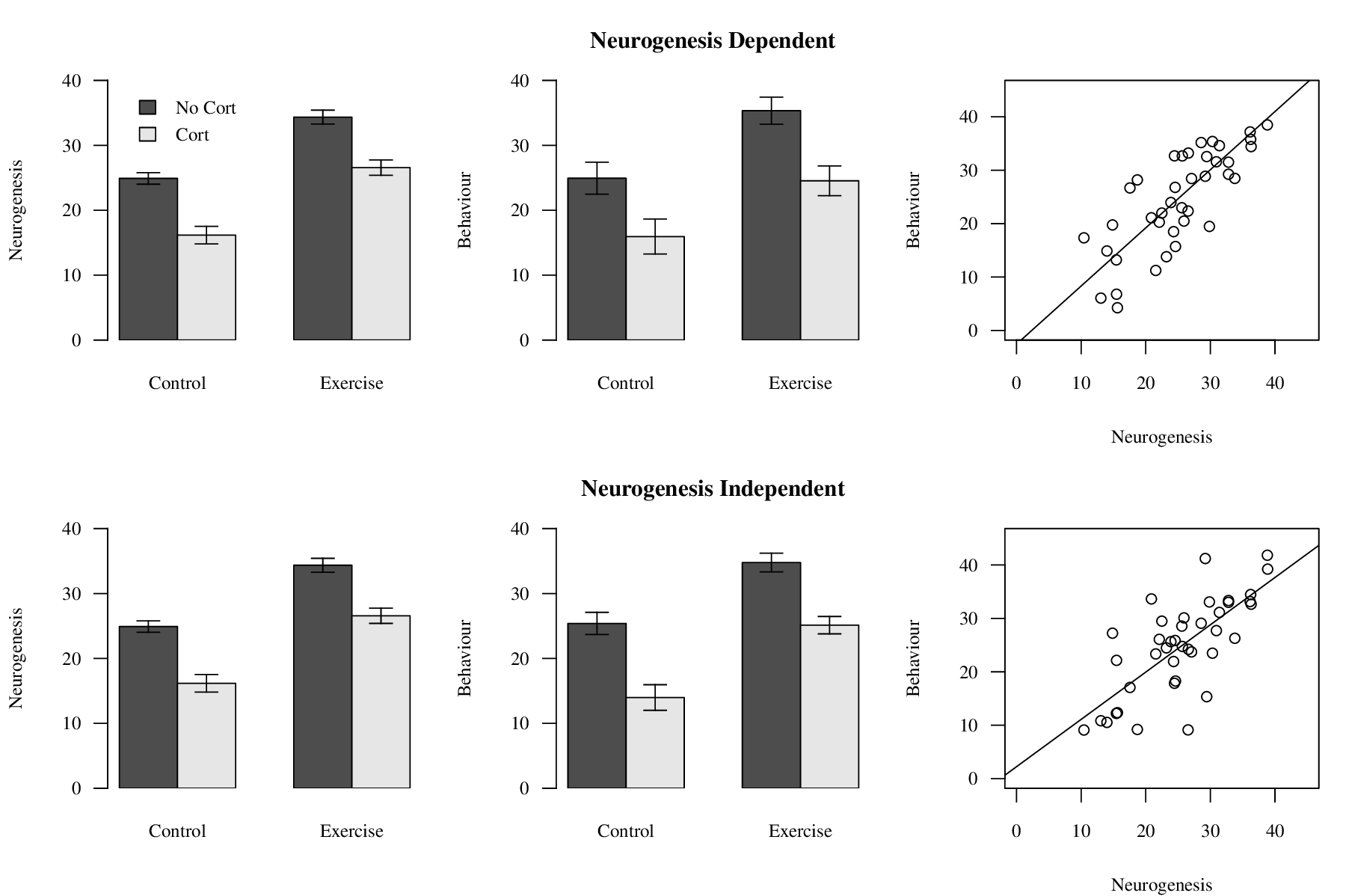}
  \caption{Comparison of data simulated from a neurogenesis-dependent and neurogenesis-independent model. These graphs and standard analyses are not able to distinguish between these models. Bars represent means $\pm$1 SEM. Cort = Corticosterone}
\end{figure}

A remarkable feature is that the data from the neurogenesis-dependent and neurogenesis-independent models are indistinguishable. The main question is whether the standard analysis of the neurogenesis-independent data will be able to conclude that neurogenesis is not important for behaviour.

\subsection*{Standard analysis of the neurogenesis-independent model data}
This data can be analysed with a 2-way ANOVA, with Exercise and Corticosterone as factors (which is also another example of a linear model), and followed by posthoc tests (Tukey's HSD), where all possible comparisons are done. This is not the recommended analysis, but it is commonly used, familiar, and thus useful for pedagogical purposes. A typical analysis and interpretation might proceed as follows: 

\begin{center}
\begin{minipage}[l]{0.8\textwidth}
Exercise increased neurogenesis (Control/No Cort vs. Exercise/No Cort, p$<$0.001), which was blocked by the addition of corticosterone (Exercise/No Cort vs. Exercise/Cort, p$<$0.001). In addition, performance on the behavioural task was significantly improved by exercise (Control/No Cort vs. Exercise/No Cort, p$<$0.001), and this effect was blocked by corticosterone treatment (Exercise/No Cort vs. Exercise/Cort, p$<$0.001). Furthermore, there was a strong correlation between neurogenesis and behaviour (Pearson correlation = 0.731, p$<$0.001).\\

Exercise increased neurogenesis and improved behavioural performance. By blocking the exercise-induced increase in neurogenesis with corticosterone, we were able to show that behavioural performance did not improve. Thus these results provide strong evidence that neurogenesis is the critical factor mediating the effect of exercise on behaviour. This is further supported by the strong neurogenesis-behaviour correlation.  
\end{minipage}
\end{center}

Recall that this analysis was based on data that was specifically created such that neurogenesis was not directly associated with behaviour ($N$ does not appear in Equation 2). This correlation was induced because neurogenesis and behaviour share a common cause in both exercise and corticosterone. Thus, neither the group comparisons nor the neurogenesis-behaviour correlation can provide evidence of causality. Unfortunately, few analyses in the neurogenesis literature proceed further. If the data were actually generated from the neurogenesis-dependent model, then the above conclusions would be correct, but this would be based on luck and not a logical consequence of the analysis. In a real experiment, it is possible that the treatments affect behaviour partly through neurogenesis and partly through other mechanisms. If this is the case, the above analysis is not able to determine the \textit{extent to which} neurogenesis is involved, which could range from 0 to 100 percent. If the aim of a study is to determine the role of neurogenesis on behaviour, then not being able to estimate how much of the observed effect is due to neurogenesis is a serious limitation of the analysis. If the relevance of neurogenesis to a particular behavioural task could be anything, the reported results have not increased our understanding.

\subsection*{A causal modelling approach}
Both models in Supplementary Figure 1 were fit to the same data as the above analysis (i.e. data for which neurogenesis was not influencing behaviour). The difference in DIC between models was 17.1 in favour of the neurogenesis-independent model, which is overwhelming support this model. Furthermore, when a model was fit containing all hypothesised relationships (i.e. all the arrows in Figure 3 in the main text), the $Neurogenesis \rightarrow Behaviour$ link was not significant ($\gamma_3$ = 0.09, 95\% CI = -0.16 to 0.35, p = 0.370), and thus the correct conclusion was reached that there is little evidence for a causal $Neurogenesis \rightarrow Behaviour$ relationship.

\section*{Conclusion}
Standard methods of analysis cannot distinguish between a situation where the only influence on behaviour is neurogenesis from a situation where the effect of neurogenesis is zero. The only question is how likely is it that neurogenesis-independent mechanisms are operating in any given study. In the main manuscript it is argued that there are known off-target effects for the commonly used methods of manipulating neurogenesis. As new methods develop, there will likely be better experimental control of the system. However, there still exists a large body of literature that is difficult to interpret.

\section*{Appendix 1: Data and R code}
\begin{verbatim}
## R code for Lazic SE (2011). Using causal models to distinguish
## between neurogenesis-dependent and -independent effects on
## behaviour (main text)



# Read in reversal and neurogenesis data
d <- data.frame(reversals=c(0, 0.512, 0.512, 0.512, 0.512, 0.512,
                  0.512, 1.488, 1.488, 0.512, 1, 1, 1.488, 1.488,
                  1.488, 2, 2, 2, 1.976, 2.976),
                neurogen=c(2.014,
                  4.892, 3.597, 2.95, 2.806, 2.23, 2.086, 2.95, 1.223,
                  5.468, 9.424, 5.036, 5.827, 4.82, 4.46, 4.892,
                  5.612, 6.691, 7.194, 6.331),
                condition=rep(0:1, times=c(9,11)) # 0 = control, 1 = runner
                )


# replicate previous results (calculate R-square)
cor.test(~reversals+neurogen, data=d)$estimate^2



# prediction 1: neurogenesis differs by condition 
LM1 <- lm(neurogen~condition, data=d) # Equation 1
summary(LM1) # prints results
confint(LM1) # calculate 95% CI

# identical analysis using a t-test, only the sign of the t-statistic
# is reversed (which is arbitrary)
t.test(neurogen~condition, data=d, var.equal=TRUE) 



# prediction 2: behaviour differs by condition
LM2 <- lm(reversals~condition, data=d) # Equation 2
summary(LM2)
confint(LM2)

# identical analysis using a t-test
t.test(reversals~condition, data=d, var.equal=TRUE)



# prediction 3: relationship between neurogenesis and behaviour
LM3 <- lm(reversals~neurogen, data=d) # Equation 3 (regression)
summary(LM3)
confint(LM3) 



# predictions 4 and 5 (different between models)
LM4 <- lm(reversals~neurogen+condition, data=d) # Equation 4
summary(LM4)
confint(LM4)




# -------------------------------------------------
# Check  assumptions


# 1) test for homogeneity of regression (i.e there should be no
# interaction between neurogenesis and condition)
LM5 <- lm(reversals~neurogen*condition, data=d)
summary(LM5)
# Interaction effect is not significant (p = 0.69)


# 2) Residuals from LM1 and LM4 should not be correlated
cor.test(resid(LM1), resid(LM4)) # p = 0.999


# 3) Neurogenesis should not be correlated with residuals from LM4
cor.test(d$neurogen, resid(LM4)) # p = 0.999





# -------------------------------------------------
# Bayesian analysis

library(R2WinBUGS)


# store data as a list
data <- list(n=20,
             neurogen = d$neurogen,
             behave = d$reversals,
             cond = d$condition 
             )


# specify the model
model <- function(){
  # cond[i] = runner/control
  # neurogen[i] = neurogenesis
  # behave[i] = behaviour (reversal data)

  for(i in 1:n)
    {
      # effect of exercise on neurogenesis (B1). Corresponds to Eq. 1 in the
      # manuscript
      neurogen[i] ~ dnorm(mean.neurogen[i], s2.neurogen)
      mean.neurogen[i] <- a1 + B1*cond[i]
      
      # effect of neurogenesis on behaviour adjusted for treatment (B5) and
      # effect of treatment on behaviour adjusted for neurogenesis (B4)
      # Corresponds to Eq. 4 in the manuscript
      behave[i] ~ dnorm(mean.behave[i], s2.behave)
      mean.behave[i] <- a2 + B4*cond[i] + B5*neurogen[i]

    }

  # priors
  a1 ~ dnorm(0, 1.0E-5) 
  a2 ~ dnorm(0, 1.0E-5)
  B1 ~ dnorm(0, 1.0E-5)
  B4 ~ dnorm(0, 1.0E-5)
  B5 ~ dnorm(0, 1.0E-5)

  s2.neurogen ~ dgamma(0.001, 0.001)
  s2.behave ~ dgamma(0.001, 0.001)

  
  # mediated effect 
  med.eff <- B1*B5
}


# save model to external file
write.model(model, "model.txt")


# Parameters to save
parms <- c("a1","a2","B1","B4","B5","s2.neurogen","s2.behave","med.eff")


# initial values for MCMC sampling
inits <- list(
         list(a1=rnorm(1), a2=rnorm(1), B1=rnorm(1), B4=rnorm(1,0,0.5),
              B5=rnorm(1), s2.neurogen=rnorm(1,1,0.05), s2.behave=rnorm(1,1,0.05)),
         list(a1=rnorm(1), a2=rnorm(1), B1=rnorm(1), B4=rnorm(1,0,0.5),
              B5=rnorm(1), s2.neurogen=rnorm(1,1,0.05), s2.behave=rnorm(1,1,0.05)),
         list(a1=rnorm(1), a2=rnorm(1), B1=rnorm(1), B4=rnorm(1,0,0.5),
              B5=rnorm(1), s2.neurogen=rnorm(1,1,0.05), s2.behave=rnorm(1,1,0.05))
              )

# run analysis
m1 <- openbugs(model.file="model.txt", data=data, inits=inits,
               parameters.to.save=parms,
               n.chains=3,  n.iter = 500000, n.thin=10)

# print results
print(m1)
m1$summary


#                 mean     sd    2.5%      25%      50%      75%    97.5%  Rhat n.eff
# a1            2.7462 0.4485  1.8588   2.4554   2.7466   3.0383   3.6265 1.001 75000
# a2            0.7568 0.3964 -0.0224   0.4990   0.7557   1.0127   1.5396 1.001 75000
# B1            3.2312 0.6026  2.0410   2.8374   3.2314   3.6211   4.4266 1.001 75000
# B4            1.0576 0.4874  0.1012   0.7415   1.0561   1.3720   2.0368 1.001 75000
# B5           -0.0309 0.1206 -0.2704  -0.1089  -0.0305   0.0476   0.2064 1.001 69000
# s2.neurogen   0.6188 0.2069  0.2815   0.4698   0.5954   0.7438   1.0839 1.001 75000
# s2.behave     2.6714 0.9187  1.1892   2.0110   2.5675   3.2178   4.7610 1.001 75000
# med.eff      -0.0995 0.3966 -0.9028  -0.3437  -0.0947   0.1494   0.6833 1.001 70000
# deviance    105.7760 4.1243 99.9155 102.7592 105.0595 108.0029 115.7389 1.001 75000


# probablility direct effect (p = 0.016)
sum(m1$sims.list$B4 < 0 ) / length(m1$sims.list$B4)
# note that the probability of the direct effect is actually 1 - 0.016 = 0.984,
# meaning that there is a high probability of the direct effect. 1 - 0.984 = 0.016
# was reported instead to keep with the convention of a small p-value indicating
# the presence of a "significant" effect.

# probability indirect effect (neurogenesis; p = 0.606)
sum(m1$sims.list$med.eff < 0 ) / length(m1$sims.list$med.eff)
\end{verbatim}

\end{document}